# Kernel-CF: Collaborative filtering done right with social network analysis and kernel smoothing

Hao Wang
Ratidar.com, Beijing, China
*haow85@live.com

## Abstract

Collaborative filtering is the simplest but oldest machine learning algorithm in the field of recommender systems. In spite of its long history, it remains a discussion topic in research venues. Usually people use users/items whose similarity scores with the target customer greater than 0 to compute the algorithms. However, this might not be the optimal solution after careful scrutiny. In this paper, we transform the recommender system input data into a 2-D social network, and apply kernel smoothing to compute preferences for unknown values in the user item rating matrix. We unifies the theoretical framework of recommender system and non-parametric statistics and provides an algorithmic procedure with optimal parameter selection method to achieve the goal.

## 1 Introduction

Recommender system is a flourishing technology that has deep impact on our daily lives thanks to its wide applications in companies such as Amazon and TikTok. Voluminous literature has been published on this technology, and a huge amount of investment has been made to commercialize the idea. The basic idea behind recommender system is to use data mining algorithms to recommend items to users based on their past preferences and historic data.

Collaborative filtering is one of the oldest recommender system technologies that is still discussed at research venues. Minor modifications have persistently appeared here and there in research publications even though most companies in the world have adopted deep learning models in their online systems.

The reason why collaborative filtering is still alive is due to its simplicity and explainability. The user-based collaborative filtering model uses the similarity between a user and others with similar preferences as weights for items not appearing in this user's historic data, and recommend items with highest preference scores to the user. The philosophy of item-based collaborative filtering is quite similar, so we omit the discussion here.

Collaborative filtering can be used as a stand-alone model, but it can also be used as the first candidate selection step for a hybrid model which uses ranking approaches after the algorithm. Used as the candidate selection model, collaborative filtering can heavily reduce the number of candidate items needed to be processed in the following ranking stage, and enabling the whole system to deal with a much larger dataset with better accuracy than stand-alone models.

In the similarity computation step of user-based collaborative filtering, the pool of items is selected by examining the positivity of similarity score. However, this practice might not be the optimal choice in theory. A smaller than necessary candidate size might cause sparsity problem in addition to poorer accuracy score. On the other hand, if the candidate size is too large, noise will be brought into the system, also leading to inferior algorithmic performance. The analysis of item-based collaborative filtering is analogous.

In recent years, there have been efforts on combining deep learning and collaborative filtering, among other modifications to the algorithm, which we will elaborate in the next section. Researchers have used statistical concepts and variants to improve the shallow model as well. However, these modifications are nearly all mild modifications.

In our research, we've found the collaborative filtering method can be generalized to something bigger than its current formulation. Therefore we aim to provide a more holistic theory assisted by social network analysis theory and non-parametric statistics (esp. Kernel smoothing). We show detailed analysis in our work, and prove by theory to the audience that a unified theoretical framework between collaborative filtering and non-parametric statistics connected by visualization.

## 2 Related Work

Collaborative filtering is a classic recommender system technology that has been used in the industry and academia for a very long time. The earliest collaborative filtering approaches are user-based collaborative filtering [1] and item-based collaborative filtering [2]. In order to make collaborative filtering more adaptive and efficient, researchers have resorted to many different approaches [3] [4] that improve the overall performance of the algorithm.

In the age of deep learning [5] [6] , collaborative filtering may seem a bit of out-of-dated. However, there are many basic questions regarding the technology that remains untouched. In 2018, Wang et.al. [7] analyzed the popularity bias of collaborative filtering and quantified the effect for the first time in history. Other questions such as the optimal neighborhood selection and unification of collaborative filtering framework with other theories are still open, which will be answered in this paper in the following sections.

One of the intermediate steps of the approach introduced in our paper is graph visualization. Popular softwares such as Gephi incoporate many visualization functionalities in the software tool itself. Algorithms such as Force Atlas 2 [8] , etc. have been widely applied in the graph visualization field.

Another important technique used in this paper is non-parametric statistics [9] [10]. To be more specific, we apply kernel smoothing approaches to redefine collaborative filtering and select the optimal neighborhood for collaborative filtering.

## 3 Collaborative Filtering

There are two types of collaborative filtering : user-based collaborative filtering (user-CF) and item-based collaborative filtering (item-CF). User-CF computes the similarity among users by comparing their historic data, and uses a weighted average formula to recommend items to users. The formal definition of user-CF is as follows :

$$R_{i,j} = \frac{\sum_{k=1}^{N} sim(user_i, user_k) R_{k,j}}{\sum_{k=1}^{N} sim(user_i, user_k)} \quad (1)$$

The key to compute the values of $R_{i,j}$ is the knowledge of $sim(user_i, user_k)$. Common practice to compute the similarity values include Jaccard distance and cosine distance, among various choices.

It is obvious from Formula (1) that the positivity of the similarity values serves as a threshold for $R_{i,k}$. A question that is not easily noticed is that whether such a threshold is proper for the user-CF algorithm. Improper selection of threshold leads to underfitting or overfitting problems.

Just like user-CF, item-CF also needs to compute the similarity values. In contrast to user-CF, item-CF focuses on the similarity between item data. The formal definition of the algorithm is as follows:

$$R_{i,j} = \frac{\sum_{k=1}^{N} sim(item_i, item_j) R_{i,k}}{\sum_{k=1}^{N} sim(item_i, item_j)} \quad (2)$$

Once again, common practices to compute the similarity values between items in item pairs are Jaccard distance and cosine distance. Not surprisingly, item-CF also needs to deal with the problem of overfitting / underfitting.

Is it possible to define CF algorithms as kernel smoothing problems ? The formulas look like Nadaraya-Watson estimator anyway. The answer is we need to do some transformation to put the user item rating matrix grid into another space if we want to transform the CF algorithms into a kernel smoothing problem.

The original user item rating grid defined by (i, j) pair where i is the user id and j is the item id with (i,j) ->$R_{i,j}$ as the scalar function defined on the grid can not be smoothed, because the neighborhood of (i,j), namely (i+1,j), (j+1,i), etc. is not the true neighborhood of (i,j) in the CF algorithms. To solve this problem, we resort to social network visualization algorithms to transform the user item rating grid into another space where the neighborhood of (i, j) is the neighborhood as defined in the CF formulas.

## 4 Social Network Visualization

Social network visualization is commercialized technology. Famous social network analysis softwares such as Gephi provides multiple choices of visualization techniques for the users. In this paper, we use ForceAtlas 2 to visualize our user-user similarity matrix in the user-CF. By effectively visualizing the similarity matrix, we are able to acquire the point coordinates of users, using which we could compute the true neighborhood of users. Not surprisingly, using the new coordinate system, CF algorithms are just a special type of Nadaraya-Watson estimator. The new coordinate system enables us to use nonparametric statistical methods other than Nadaraya-Watson estimator.

The classic Force Atlas 2 algorithm considers the visualization layout of social network data as the interaction between 2 forces, the attraction force and the repulsion force. The attraction force is defined as follows :

$$F_a(n_1, n_2) = d(n_1, n_2) \quad (3)$$

The repulsion force is defined below :

$$F_r(n_1, n_2) = k_r \frac{(deg(n_1)+1)(deg(n_2)+1)}{d(n_1, n_2)} \quad (4)$$

Variables in Formula (3) an (4) are defined as follows: $n_1$ and $n_2$ represent users, d($n_1, n_2$) is the distance and deg() is the degree of the user in the social network graph. The algorithm is inspired by the real-world physical phenomenon of interactions among electrically charged particles.

Please notice we could also visualize the recommender system similarity matrix using Multidimensional Scaling or Isomap algorithms. These 2 algorithms preserve the distance / similarity among social network nodes in lower dimensions. However, they are too much slower than Force Atlas 2, and it would be impossible to run the algorithms on hundreds of millions of users in real commercial environments in big internet corporations.

Fig.1 showcases a snapshot of a ForceAtlas 2 simulation on LDOS-CoMoDa dataset :

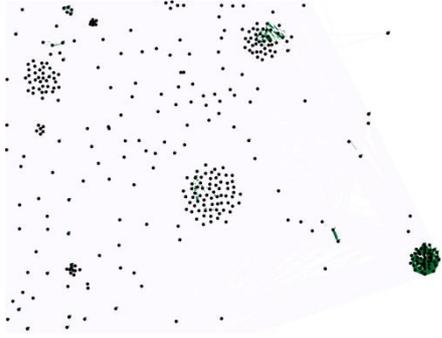

Fig. 1 Force Atlas 2 simulation on LDOS-CoMoDa dataset

From Fig.1, we observe clustering properties of the users in the user-user similarity graph. Similar users are clustered together in the visualization while non-similar users are separated apart from each other. The neighboring points (users) in a cluster naturally form the neighborhood of a point (user) in the graph. CF algorithms are simply Nadaraya-Watson estimators in the visualization graph, which serves as a new coordinate space transformed from the original user item rating matrix grid.

## 5 Kernel Smoothing

In the previous section, to compute the visualization graph of user-user / item-item similarity, we need the distance between 2 points in the graph. However, what we have at hand is the similarity data. We adopt the common practice of taking the inverse of similarity to generate the distance :

$$d(n_1, n_2) = \frac{1}{sim(n_1, n_2)} \quad (5)$$

To build the relationship between CF algorithms and **Nadaraya-Watson** estimator, we review the definition of Nadaraya-Watson estimator at first :

$$\hat{m} = \frac{\sum_{i=1}^{n} K_{h_x}(x - x_i) y_i}{\sum_{i=1}^{n} K_{h_x}(x - x_i)} \quad (6)$$

If we replace the kernels $K_{h_x}(x - x_i)$ with $sim(x, x_i)$, we obtain the CF algorithms. The most direct replacement would be similarity computed using Euclidean distance or Jaccard distance on implicit feedback.

By the above reasoning, if we transform the original user item rating matrix grid into the social network graph coordinate system, and use the new coordinates of user-user / item-item pair, we successfully reduce the original CF algorithms into Nadaraya-Watson estimation problem.

There exists a remaining problem in our reduction framework - how to select the optimal neighborhood size for the Nadaraya-Watson kernel. In the language of CF algorithms, the problem is equivalent to how to choose the best number of users / items used in the similarity computation steps.

The optimal neignborhood size for Nadaraya-Watson estimator kernel in 1-D domain is computed below :

$$h_* = \left(\frac{1}{n}\right)^{1/5} \left(\frac{\sigma^2 \int K^2(x)dx \int dx/f(x)}{(\int x^2 K^2(x)dx)^2 \int \left(r''(x) + 2r'(x)\frac{f'(x)}{f(x)}\right)^2 dx}\right)^{1/5} \quad (7)$$

In Formula (7), $h_*$ is the optimal bandwidth for Nadaraya-Watson estimator; K is the kernel function; x represents the user or item; f() is the probabilistic distribution function from which x's are sampled; r is the function defined as follows :

$$r(x) = \sum_{i=1}^{n} K_{h_x}(x - x_i) y_i \quad (8)$$

The algorithm needs estimation of the probabilistic density function f(), which can be computed using kernel density estimation.

We now generalize the 1-D optimal bandwidth result to 2-D domain, which is the dimension of the social network graph - our new coordinate system for CF algorithms. In 2-D we have 2 optimal bandwidths $b_u$ and $b_t$ to compute :

$$b_t = \left(\frac{\sigma^2 R(K)^2 I_{uu}^{3/4} I_f}{\mu_2(K)^2 I_{tt}^{3/4} \left(I_{tt}^{1/2} I_{uu}^{1/2} + I_{tu}\right) n}\right)^{1/6} \quad (9)$$

$$b_u = (I_{tt}/I_{uu})^{1/4} b_t \quad (10)$$

The variables in Formula (9) and (10) are defined as follows :

$$I_{tt} = \int_A \omega(t, u) r^{(2,0)}(t, u)^2 d(t, u) \quad (11)$$

$$I_{uu} = \int_A \omega(t, u) r^{(0,2)}(t, u)^2 d(t, u) \quad (12)$$

$$I_{uu} = \int_A \omega(t, u) r^{(2,0)}(t, u) r^{(0,2)}(t, u) d(t, u) \quad (13)$$

$$I_f = \int_A \omega(t, u) f(t, u)^{-1} d(t, u) \quad (14)$$

$$R(K) = \int_{-1}^{1} K(x)^2 dx \quad (15)$$

Variable r represents the underlying function of the point data in 2-D domain. The Nadaraya-Watson estimation in 2-D domain is formally defined in the following way :

$$r(t, u; b_t, b_u) = b_t^{-1} b_u^{-1} \sum_{i=1}^{n} \int_{A_i} K\left(\frac{t-v}{b_t}\right) K\left(\frac{u-w}{b_u}\right) d(v, w) Y_i \quad (16)$$

The variable $\omega$ is a function introduced intentionally to confine the functionals on the region A. $r^{(2,0)}$ and $r^{(0,2)}$ are second order partial derivatives of r. We assume the social network vertices is disturbed in the following way :

$$Y_i = r(t_i, u_i) + \varepsilon_i \quad (17)$$

Function f is defined as the probability density function of data noise $\varepsilon_i$.

After the mathematical derivation of the optimal bandwidth / neighborhood selection, we need to put the theory into practice. In other words, we need to find out ways to approximate the unknown values in Formula (9) and (10). To approximate the partial derivatives of r, we use ordinary least squares to compute the underlying function r and the derivatives. To estimate the density function f, we use Kernel Density Estimation :

$$f(\mathbf{x}, \mathbf{H}) = \frac{1}{n}\sum_{i=1}^{n} K_{\mathbf{H}}(\mathbf{x} - \mathbf{X_i}) \qquad (18)$$

We apply the **Reference Rule** to compute the optimal value of **H** as follows :

$$H_{REF} = \left(\frac{1}{n}\right)^{1/3} \hat{\Sigma} \qquad (19)$$

, where $\hat{\Sigma}$ is the empirical covariance matrix :

$$\hat{\Sigma} = \frac{1}{n-1}\sum_{i=1}^{n}(X_i - \bar{X})(X_i - \bar{X})^T, \text{ with } \bar{X} = \frac{1}{n}\sum_{i=1}^{n} X_i$$

Please notice our reduction of CF algorithms to kernel smoothing problem can be generalized to kernel smoothing with polynomial degree greater than 0, such as local weighted linear regression. Likewise, optimal neighborhood size selection problem can be solved using either plug-in method or cross-validation method, as discussed in [9] and [10].

We name our new non-parametric approach in replacement of CF algorithms Kernel CF. The formal procedure is actually an online learning algorithm (We only illustrate the procedure for User-based Kernel CF. Item-based Kernel CF is executed similarly.):

1. Compute the neighborhood of each user using classic user-based collaborative filtering.
2. Transform the existing user item rating matrix grid into a new coordinate system defined by Force Atlas 2 algorithm.
3. For each user in the new coordinate system, compute the optimal neighborhood of each user using Nadaraya-Watson kernel with optimal bandwidth computed by Formula (9) and (10) with respect to the candidate list generated by the classic collaborative filtering algorithm.
4. Recommend items to the user using the new optimal neighborhood obtained in 3.

Our algorithm is actually an approach that finds the optimal candidate pool for existing collaborative filtering. Originally, the task is difficult to achieve because researchers could not find a proper coordinate system in which non-parametric statistical approaches such as plug-in method can be applied.

# 6 Conclusion

In this paper, we unify the classic collaborative filtering with non-parametric statistics and provide the formulas to compute the optimal neighborhood. The algorithm requires two-steps: 1. Computation of candidate pool using the classic Collaborative Filtering approach. 2. Re-select the candidate pool using non-parametric statistical tools as discussed in this paper.

This is the first time in AI history that a practical AI problem is solved via the bridge between 3 different scientific domains, namely recommender system, information visualization, and non-parametric statistics. We hope more researchers are inspired to explore multi-disciplinary science after reading our paper.

In future work, we would like to examine other recommendation algorithms such as matrix factorization, factorization machines, and deep learning, to see if there is unified theory underlying all of the afore-mentioned technologies.